\RequirePackage{fix-cm}
\documentclass[twocolumn]{svjour3}          
\smartqed  
\usepackage{graphicx}
 \usepackage{mathptmx}      
\usepackage{latexsym}
\usepackage{bm}
\usepackage{amsmath}
\usepackage{cases}
\begin{document}

\title{Fine Particles as Solute in Fine Particle (Dusty) Plasmas}
\subtitle{Phase Separation and Diagram}
\author{Hiroo Totsuji}
\institute{H. Totsuji \at
              Okayama University (Prof. Emeritus), Okayama, Japan \\
              \email{totsuji-09@t.okadai.jp}            \\
             \emph{Present address: Saginomiya 6-3-16, Nakanoku, Tokyo, Japan}  
}

\date{Received: date / Accepted: date}

\maketitle

\begin{abstract}
Analysis has been made on a possibility of phase separation and critical point 
as a solution of fine particles in plasmas,
extending the previous work 
which pointed out this as one of two such possibilities.
Fine particles are treated as a solute in the solvent of weakly coupled plasmas
of electrons and ions.
Examples of phase diagram have been obtained
and
coexisting phases are shown to have fine particles
which differ in densities by almost one order of magnitude.
\keywords{Fine particle plasma 
\and Phase separation as solution \and Phase diagram \and Critical point}
\end{abstract}

\section{Introduction}
\label{intro}
Fine particle (dusty) plasmas are charge-neutral mixtures
of negatively charged fine (dust) particles and weakly ionized background plasmas
(electrons, ions, and neutral gas atoms).
Since particles in fine particle plasmas have negative charges of large magnitude,
they are often in the state of strong coupling.
Various interesting phenomena related to strongly coupled fine particles
are expected\cite{SM02}
and 
have been observed
especially in microgravity experiments\cite{FMP05,TMF08}.
We here assume usual experimental conditions,
for example,
gas pressure of $10-10^2\ {\rm Pa}$
and
electron and ion densities of $10^8 - 10^9\ {\rm cm^{-3}}$
and
refer to fine particles simply as `particles'.

Based on drift-diffusion equations,
it has been shown\cite{HT16a,HT18} that,
both under microgravity and gravity,
the existence of particles largely enhances the charge neutrality 
while
{\it affecting the electron density very little};
Particles seem to accompany charge-compensating (extra) ions.

The mean free path of electrons 
is usually comparable or larger than the typical size $(\le 1\ {\rm cm})$ of the domain of our interest.
For example,
in the Ar gas of pressure $p_n$ and temperature $T_n$
with the neutral atom density $n_n$
\begin{align*}
n_n \sim 2.4\cdot 10^{14}\ p_n{\rm [Pa]}\left({300\ {\rm K} \over T_n }\right){\rm cm^{-3}},
\end{align*}
the electron-Ar collision cross section $\sigma_{en} \sim 2\cdot 10^{-16}\ {\rm cm^2}$ gives the mean-free path
\begin{align*}
{1 \over n_n \sigma_{en}} 
\sim 2.1 \cdot 10\ {T_n /300\ {\rm K} \over p_n\ {\rm [Pa]}}\ {\rm cm}.
\end{align*}
When $p_n \sim 10-10^2\ {\rm Pa}$,
the electron mean free path is thus $\sim 0.2-2\ {\rm cm}$.
This is in contrast with the ion mean free path:
The ${\rm Ar}^+-{\rm Ar}$ collision cross section $\sigma_{in} \sim 1\cdot 10^{-14}\ {\rm cm^2}$ gives 
the ion mean free path much smaller than that of electrons 
by a factor $\sim 50$.
These numbers are consistent with the insensitivity of electron density to the particle density
changing in scales smaller than 1 cm.

There have been pointed out two possibilities of critical phenomena\cite{HT11},
both coming from the strong coupling of particles.
One (case 1) is a thermodynamic instability of the whole system
leading to coexisting phases with different electron densities.
Since the ideal gas contribution of electrons to the pressure
is much larger than those of ions and particles,
this requires very strong coupling between particles
which gives comparable negative contribution to the pressure.
The other (case 2) is in the domain where
electrons still determine the total pressure
and
coexisting phases have the same electron density
with different particle densities:
Particles are regarded as a solute in the solvent of electron-ion plasma.
In view of the insensitivity of the electron density 
to particle density in particle clouds,
we here extend the analysis of the latter possibility.

\section{Condition for coexisting phases and phase diagram}
\label{sec:1}

We adopt the same notations as in Ref.6:
Charges, densities and temperatures of electrons, ions, and fine particles
are expressed by $(-e, n_e, T_e)$, $(e, n_i, T_i)$ and $(-Qe, n_p, T_p)$, respectively,
and their values are similarly assumed as
$n_e \sim n_i \sim (10^8-10^9)\ {\rm cm^{-3}}$, $n_p \sim 10^5\ {\rm cm^{-3}}$, 
$k_B T_e \sim {\rm a\ few\ eV}$, and $T_i \sim T_p \sim 300\ {\rm K}$.
The background plasma (electrons and ions) is weakly coupled
and
particles interact via the Yukawa potential
\begin{align*}
{(Qe)^2 \over 4 \pi \varepsilon_0 r}\exp \left(-{r \over \lambda}\right)
\end{align*}
with the screening length $\lambda$ determined mainly by ions
\begin{align*}
\lambda \sim \left({\varepsilon_0 k_B T_i \over e^2 n_i}\right)^{1/2}
= 120\ \left[ \left({T_i \over 300\ {\rm K}}\right)\left( {10^8\ {\rm cm^{-3}} \over n_i}\right)\right]^{1/2} {\rm \mu m}.
\end{align*}
The Yukawa system of particles is characterized 
by the coupling parameter $\Gamma$ and the strength of screening $\xi$
defined respectively by the mean distance $a$ as
\begin{align*}
\Gamma = {(Qe)^2 \over 4 \pi \varepsilon_0 a k_B T_p},
\ \ \ \xi={a \over \lambda},
\ \ \ a=\left({3 \over 4 \pi n_p}\right)^{1/3}.
\end{align*}
Since $Q$ takes on values $\sim 10^3$, $\Gamma$ can be very large.

We write the Helmholtz free energy of the whole system $F$ in the volume 
$V\ (N_{e, i, p}=n_{e, i, p}V)$ in the form
\begin{align}
F
=
&F^{(e)}_{\rm id}(T_e, V, N_e) + F^{(i)}_{\rm id}(T_i, V, N_i) + F^{(p)}_{\rm id}(T_p, V, N_p) \nonumber \\
&+ N_p k_B T_p f^{(p)},
\end{align}
$F^{(e, i, p)}_{\rm id}$ and $N_p k_B T_pf^{(p)}$ being
ideal gas values
and
the non-ideal part of particle free energy, respectively.
Applying results of Yukawa numerical simulations
and
taking into account the finite particle radius $r_p$,
we have obtained an approximate expression\cite{HT08a} [(19) in Ref.6]
\begin{align}\label{int-Yukawa-finite-radius}
f^{(p)}(\tilde{\Gamma}, \xi, \tilde{r}_p)
&\approx
a_1 \tilde{\Gamma} \exp(a_2\xi) + 4a_3 \tilde{\Gamma}^{1/4}\exp(a_4\xi) \nonumber \\
&+{3 \over 2}\tilde{\Gamma}\xi^{-2}[1-(1+2\tilde{r}_p)\exp(-2\tilde{r}_p)] \nonumber \\
&- {1 \over 2} \tilde{\Gamma} \xi (1+\tilde{r}_p) \exp(-2\tilde{r}_p),
\end{align}
where $a_1=-0.896$, $a_2=-0.588$, $a_3=0.72$, $a_4=-0.22$,
\begin{align}
\tilde{\Gamma}=\Gamma {\exp(2\tilde{r}_p) \over (1+ \tilde{r}_p)^2},
\end{align}
and
\begin{align}
\tilde{r}_p={r_p \over \lambda}.
\end{align}
Quantities $\tilde{\Gamma}/\Gamma$ and $\tilde{r}_p$ express the effect of finite particle radius $r_p$:
When $r_p=0$, $\tilde{\Gamma} = \Gamma$ and $\tilde{r}_p=0$.

When $n_e \approx n_i \gg Q n_p$,
particles can be regarded as a solute in the solvent background plasma
and
there exists a possibility (case 2)
of coexisting phases 
with the same electron density but different particle densities\cite{HT11}.
For coexisting phases, I and II, required is the condition [(49) in Ref.6]
\begin{align}\label{phase-condition}
k_B T_p \ln {n^{I}_p \over n^{II}_p} + Q k_B T_i \ln {n^{I}_e + Qn^{I}_p \over n^{II}_e + Qn^{II}_p}
=
-\left[\Delta \mu^{(p)} \right]^{I}
+\left[\Delta \mu^{(p)} \right]^{II}.
\end{align}
Here,
$\Delta \mu ^{(p)}$ 
denotes the non-ideal part of the chemical potential of particles $\mu^{(p)}$
as [(47) in Ref.6]
\begin{align}
\mu^{(p)}=\mu_{\rm id}^{(p)} + \Delta \mu ^{(p)}
\end{align}
and
calculated from (\ref{int-Yukawa-finite-radius}) in the form\cite{footnote1} [(50) in Ref.6]
\begin{align}
{\Delta \mu^{(p)} \over k_B T_p}
&=
{1 \over 3}a_1 \tilde{\Gamma} \exp(a_2\xi)(4-a_2\xi) \nonumber \\
&+ {1 \over 3} a_3 \tilde{\Gamma}^{1/4}\exp(a_4\xi)(13 - 4 a_4\xi) \nonumber \\
&+ 3 \tilde{\Gamma}\xi^{-2}[1-(1+2\tilde{r}_p)\exp(-2\tilde{r}_p)] \nonumber \\
&- {1 \over 2} \tilde{\Gamma} \xi (1+\tilde{r}_p) \exp(-2\tilde{r}_p).             
\end{align}
The critical condition for the appearance of coexisting phases
is written in the usual form\cite{LLSP} as
\begin{align}
{\partial \tilde{\mu} \over \partial n_p}=0
\end{align}
for the effective chemical potential $\tilde{\mu}$
\begin{align}\label{effective-c.p.}
\tilde{\mu}
=k_B T_p \ln n_p + Q k_B T_i \ln (n_e + Qn_p) + \Delta \mu^{(p)},
\end{align} 
the second term coming from the charge neutrality condition.

The coexisting phases are determined by the relation (\ref{phase-condition}).
When we change $n_p$ (keeping $n_e$ and $n_i$ unchanged),
the system follows the curve given by [(51) in Ref.6]
\begin{align}
\Gamma \xi \approx {(4\pi n_e)^{1/2} Q^2e^3 \over k_B T_p(k_B T_i)^{1/2}} 
= {\rm const}.
\end{align}
Estimating the value of $Q$ by\cite{footnote2}
\begin{align}
Q \sim 0.5 {k_B T_e \over e^2 /4 \pi \varepsilon_0 r_p}
\sim 3.5 \cdot 10^2  (k_BT_e[{\rm eV}]) r_p[{\rm \mu m}],
\end{align}
we have
\begin{align}
&{(4\pi n_e)^{1/2} Q^2e^3 \over k_B T_p(k_B T_i)^{1/2}} \nonumber \\
&\sim 5.5 \cdot 10  {(k_BT_e[{\rm eV}])^2 (r_p[{\rm \mu m}])^2
\left({n_e / 10^8\ {\rm cm^{-3}}}\right)^{1/2}
\over 
\left(T_p / 300\ {\rm K} \right)
\left(T_i / 300\ {\rm K} \right)^{1/2}}.
\end{align}
Note that the value of $\Gamma \xi$ covers a wide range;
When $n_e \sim 10^8\ {\rm cm^{-3}}$
with $k_B T_e \sim (1-3)\ {\rm eV}$
and $r_p \sim (1-3)\ {\rm \mu m}$,
$\Gamma \xi \sim (5.5 \cdot 10 -4.5 \cdot 10^3)$
and
when $n_e \sim 8\cdot 10^8\ {\rm cm^{-3}}$,
$\Gamma \xi \sim (1.6 \cdot 10^2 -1.3 \cdot 10^4)$.

An example of the behavior of $\tilde{\mu}$
on the curve $\Gamma \xi = {\rm const.}=C$ is shown in Fig.1.
With the increase of $C$,
$\tilde{\mu}$ changes from a monotone function of $\Gamma$ 
to the one with two extrema,
signaling the beginning of coexistence.
\begin{figure}
  \includegraphics[width=9cm]{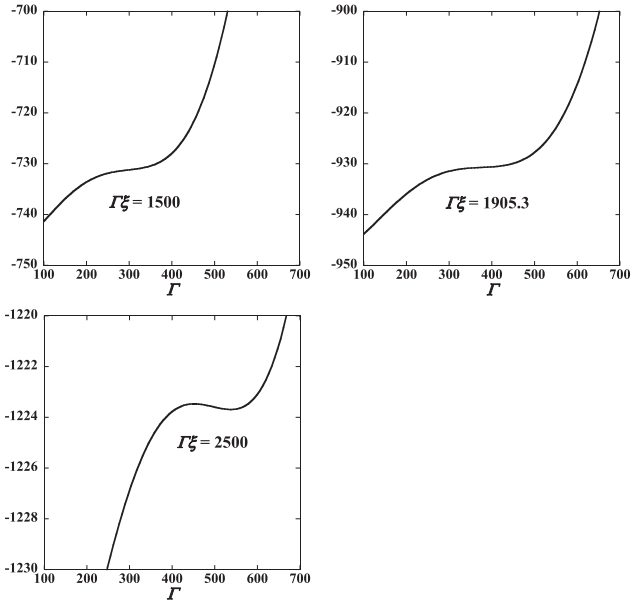}
\caption{Behavior of normalized effective chemical potential 
$\tilde{\mu}/k_B T_p$ on the curve $\Gamma \xi = C$ with $\tilde{r}_p=8.37\cdot 10^{-3}$; In this case, $(\Gamma \xi)_c=1905.3$, $\Gamma_c=373.0$, and $\xi_c=5.11$.
With the increase of $C$,
extrema appear signaling phase coexistence.
(The origin of ordinate is arbitrarily shifted.)}
\label{fig:1}
\end{figure}
The phase diagram is shown in Fig.2.
When the value of $\Gamma \xi$ is less than $(\Gamma \xi)_c$,
we have one phase irrespective of the value of $n_p$.
When $\Gamma \xi =(\Gamma \xi)_c$,
we have a critical point at $(\Gamma_c, \xi_c)$
and when $\Gamma \xi > (\Gamma \xi)_c$,
we have the domain of two coexisting  phases\cite{footnote3}.
As shown in Fig.3,
the critical value and phase boundary are dependent on the parameter 
$\tilde{r}_p = r_p / \lambda$.
The critical point $(\Gamma_c, \xi_c)$ is approximately expressed by interpolation formulae,
\begin{align}
&\Gamma_c \approx 3.74\cdot 10^2-94.5\ \tilde{r}_p+5.63 \cdot 10^3\ \tilde{r}^2_p,  \\
&\xi_c \approx5.11+0.15\ \tilde{r}_p +2.3\ \tilde{r}^2_p.
\end{align}
\begin{figure}
  \includegraphics[width=8cm]{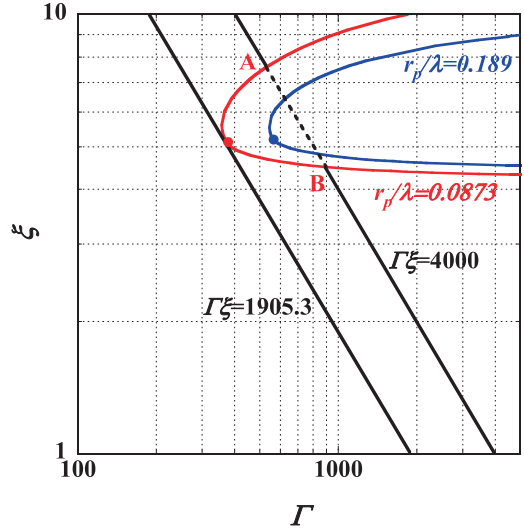}
\caption{Phase diagram 
for cases of $\tilde{r}_p=8.73\cdot 10^{-3} 
({\rm e.g.},\ r_p=1\ {\rm \mu m},\ n_e=10^8\ {\rm cm^{-3}})$
and $\tilde{r}_p=1.89\cdot 10^{-1} 
({\rm e.g.},\ r_p=8\ {\rm \mu m},\ n_e=8 \cdot 10^8\ {\rm cm^{-3}})$,
filled circles being critical points.
Two phases coexist inside the phase boundary curve.
Straight lines are examples of $\Gamma \xi = C$
with $C=1905.3$ and $4000$.
When $\Gamma$ increases along $\Gamma \xi =4000$
in the case of $\tilde{r}_p=8.73\cdot 10^{-3}$,
two phases (A and B) coexist between points A and B.}
\label{fig:2}
\end{figure}
\begin{figure}
  \includegraphics[width=8.5cm]{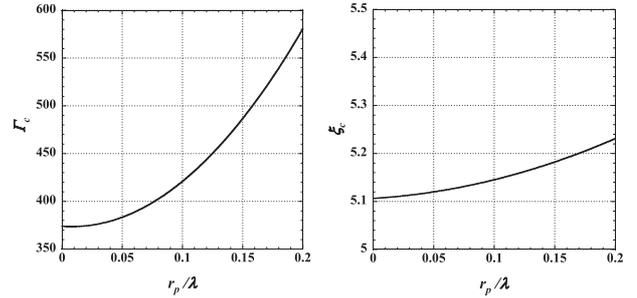}
\caption{Dependence of $\Gamma_c$ and $\xi_c$ on $\tilde{r}_p = r_p/\lambda$.}
\label{fig:3}
\end{figure}

\section{Discussions}
\label{sec:2}

Correlations between charged particles
generally give negative contribution to their internal energy 
and,
if the charge neutrality is somehow satisfied,
they tend to coagulate.
In our case,
some strength of coupling is necessary
for particles accompanying charge-compensating (extra) ions
to coagulate:
The second term in (\ref{effective-c.p.})
corresponds to this situation.

Values of $(\Gamma, \xi)$ for coexisting phases
are determined 
by the cross-section of $\Gamma \xi={\rm const.}$
and the phase boundary curve.
As shown in Fig.2,
the ratio of $\Gamma$ in coexisting phases is $\sim 2$,
if not very close to the critical point.
The ratio of the density is thus $\sim 2^3=8$.
We may therefore expect coexisting phases
with the density difference of nearly one order of magnitude.

Sharp boundaries of particle clouds or voids have been often observed
in microgravity experiments.
Since we have finite electric field and ion flow
on the periphery of particle clouds even under microgravity,
it may not be easy to separate the effect of strong coupling
on the sharpness of boundaries.
For example,
the formation of voids is usually attributed to the ion flow\cite{GMTV99,HT16a}.
We expect, however, 
detailed control of parameters might reveal 
the contribution of the strong coupling
in the formation of sharp boundaries of particle clouds.
We also point out that
the above phenomena in the horizontal structure 
could be observed in experiments under gravity,
where
detailed control of parameters is easier.


%
%


\begin{thebibliography}{}
\bibitem{footnote1}
On the left-hand side of (50) in Ref.6,
the denominator $k_B T_p$ is missing.
\bibitem{footnote2}
The value of $Q$ can be estimated, for example, 
by the orbit-motion-limited (OML) theory.
The magnitude of $Q$ is proportional to the capacitance of the particle
and a typical value of the coefficient is $\sim 0.5$.
\bibitem{footnote3}
The critical values $(\Gamma \xi)_c, \Gamma_c, \xi_c$
given in Ref.6
should be corrected as given here.
\end{thebibliography}


\end{document}